# Effects of Translation-Rotation Coupling on the Displacement Probability Distribution Functions of Boomerang Colloidal Particles

Ayan Chakrabarty,[a] Feng Wang,[a] Kai Sun,[b] Qi-Huo Wei[a]*
a. Liquid Crystal Institute, Kent State University, Kent, OH 44242
b. Department of Materials Science and Engineering, University of Michigan, Ann Arbor, MI

Prior studies have shown that low symmetry particles such as micro-boomerangs exhibit behaviour of Brownian motion rather different from that of high symmetry particles because convenient tracking points (TPs) are usually inconsistent with the center of hydrodynamic stress (CoH) where the translational and rotational motions are decoupled. In this paper we study the effects of the translation-rotation coupling on the displacement probability distribution functions (PDFs) of the boomerang colloid particles with symmetric arms. By tracking the motions of different points on the particle symmetry axis, we show that as the distance between the TP and the CoH is increased, the effects of translation-rotation coupling becomes pronounced, making the short-time 2D PDF for fixed initial orientation to change from elliptical to crescent shape and the angle averaged PDFs from ellipsoidal-particle-like PDF to a shape with a Gaussian top and long displacement tails. We also observed that at long times the PDFs revert to Gaussian. This crescent shape of 2D PDF provides a clear physical picture of the non-zero mean displacements observed in boomerangs particles.

## 1. Introduction

Brownian motion as the governing process of diffusion in a variety of physical, chemical and biological processes has been a subject of extensive research over the past century.[1–4] The physical essence of Brownian motion is the random displacements of microscopic particles under the stochastic impacts of surrounding molecules.[5] The probability distribution of the random displacements has shown to be Gaussian for spheres by Einstein and others.[1,2,5,6] Studies, however, have shown that non-Gaussian distributions of random displacements are more common than expected in soft matter systems owing to multiplicative noises of different physical origins.[7,8] The experimental and theoretical studies by Han et al on ellipsoidal particles show that the anisotropic diffusion coefficients make the random displacements non-Gaussian in the lab frame, while their mean square displacements remain linear at all times.[9] Single particle experiments by Wang et al show that Fickian while non-Gaussian Brownian motion of spherical particles can occur often as a result of coupling between the particle and its surrounding environments such as linear phospholipid tubes and entangled f-actin networks.[8,10]

While the hydrodynamic theory for the Brownian motion of particles with irregular shapes has been established by Brenner and others,[11–16] the advances in fabricating passive and active particles with well-designed geometric shapes have led to interests in the Brownian motion and active swimming behavior of particles with low symmetry shapes.[17–24] In our previous publications, we showed that the diffusion behavior of low-symmetry particles such as boomerangs is generally different from high symmetry particles like ellipsoids.[19,20] In two dimensions (2D), there always exists one unique point fixed to the particle, named the center of hydrodynamic stress (CoH) to which the translational and rotational motions are decoupled. For low symmetry particles, convenient tracking points (TPs) such as the symmetry center (if any) and the center of mass (CoM) normally do not overlap with the CoH, leading to the coupling of translational and rotational motions. As a result, the mean square displacements (MSDs) exhibit a crossover from short- to long-time diffusion with different diffusion coefficients.[19,20] Further, another interesting and rather intriguing behavior that arises as a consequence of this coupling is the non-zero mean displacement of the particles for a fixed initial orientation and the motion is biased towards the CoH. In contrast, for high symmetry particles, when the CoM is conveniently used for motion tracking, these interesting behaviors originating from the translation-rotation coupling are absent because the CoM is coincident with the CoH.[20]

As for theoretical and numerical studies, Delong et al developed an overdamped Langevin formulation and robust numerical algorithms that allow for simulations of the diffusive and active motion of rigid bodies of arbitrary shapes under various boundary conditions.[25,26] By Brownian dynamic simulations of several exemplary colloidal particles (including boomerangs) sedimented on a non-slip substrate under gravity, they verified the important effects of tracking points on diffusion behaviors and obtained agreements with the experiments.[25,26] A rigid multiblob method was developed by Usabiaga et al for numerically calculating the mobility of passive and active particles of complex shapes in confined or unconfined geometries.[27] In addition, Cichocki et al derived the analytical expressions of the time-dependent cross-correlations between the translational and rotational displacements for arbitrarily shaped particles.[28]

In this paper we focus an open while important question regarding how the translation-rotation coupling affects the probability distribution functions (PDFs) of the random displacements of the boomerang colloidal particles. By using different points fixed on the particle symmetry axis for motion tracking, we vary the translation-rotation coupling and observe that with the increase of distance between the TP and the CoH, the 2D PDF for fixed initial angle change from an elliptical shape to crescent shape. This crescent shape of the 2D PDF manifests the effects of translation and rotation coupling and provides a clear physical picture of the non-zero mean displacements observed in boomerangs particles. We also measured the angle averaged PDFs, and found that they can be scaled into different master curves at short and long times. The long time master curves can be verified to be perfectly Gaussian while the short time master curve vary with the translation-rotation coupling (i.e. the distance between the TP and CoH) which can be fitted with two different empirical formulae. Lastly, we discuss the implications of these results to prior or potential experiments with other particle shapes.

## 2. Experimental

The particles and motion trajectories used in this work are the same as those used in the prior publication.[19] For completeness, we briefly describe the experimental procedures here. The boomerang particles used in this study were made of photo-curable polymer SU8 (from MicroChem) by projection photolithography, and the details of the fabrication processes have been described in a previous publication.[29] The particles used in this study have a 2.1 μm arm length, 0.51 μm × 0.55 μm arm cross-section and 90° apex angle. Aqueous suspensions of these boomerang particles are stabilized by adding 1mM sodium dodecyl sulfate (SDS). Cells made of glass slides with ~ 2 μm gap thicknesses were used to confine the colloidal particles in a quasi-two dimensional geometry. Particle concentrations were kept very low so that there are no hydrodynamic interactions between them.

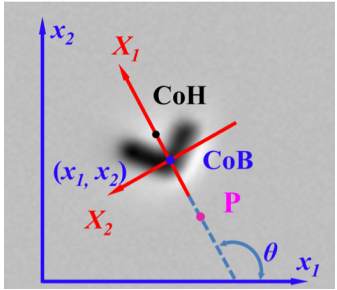

**Fig.1** Optical microscopic picture of a boomerang colloidal particle confined in a quasi-2D cell and schematics of the lab and body frame coordinate systems.

Brownian motion of the boomerangs was observed using an inverted optical microscope, and motion videos were recorded using a digital CCD camera. Each video contains 3000 frames with 0.05 seconds time interval (τ). A total of 167 videos of the same boomerang particle were recorded and analyzed by using the high accuracy image processing algorithm as described in a previous paper.[29] The cross point of the central axes of two arms, termed as the center of the body (CoB) was initially used for position tracking, and the bisector of the apex angle was used to represents the particle orientation (Fig.1). These trajectories were then merged in 24 different random sequences into a long trajectory totaling about 12 million frames.

## 3. Results

Firstly, we summarize our prior findings regarding the Brownian motion of the boomerangs with symmetric arm lengths. The rotational Brownian motion is independent of the tracking point used, and the MSDs of particle orientation angles grow linearly with time for all times. The measured rotational diffusion coefficient is $D_\theta = 0.045\ rad^2/s$. In contrast, the translational Brownian motion is highly sensitive to the point used for motion tracking. Since the convenient tracking point is generally not coincident with the CoH for the low symmetry particles like boomerangs, the MSDs for translational motions in the lab frame grow linearly with time only at short and long times with two different diffusion coefficients due to the translation-rotation coupling. The anisotropic diffusion coefficients measured in the body frame are shown to be highly dependent on the position of the TP. A Langevin model revealed the analytical relationships between the anisotropic diffusion coefficients and the diffusion coefficients measured at the CoH. Here we consider only TPs on the apex angle bisector line, and assume the distance between the CoH and the tracking point is *r*, then the anisotropic diffusion coefficients can be expressed as:[19]

$$D_{11} = D_{11}^{CoH}, D_{22} = D_{22}^{CoH} + r^2 D_\theta \quad (1a)$$
$$D_{1\theta} = 0; D_{2\theta} = r D_\theta \quad (1b)$$

When the cross point of two arms, i.e. the Center of the Body (CoB) is used for motion tracking, the measured anisotropic diffusion coefficients are $D_{11} = 0.049\ \mu m^2/s$, $D_{22} = 0.117\ \mu m^2/s$, and $D_{2\theta} = -0.051\ \mu m \cdot rad/s$. Based on these diffusion coefficients, the center of hydrodynamic stress (CoH)

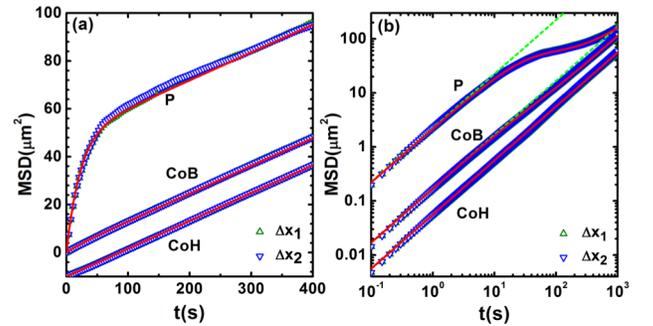

**Fig. 2** (a) Measured MSDs for the three tracking points. Red lines are the theoretical curves using Eq. 2 and $D_\theta = 0.044\ rad^2/s$, and $\overline{D}^{CoH} = 0.054\ \mu m^2/s$. (b) Data in (a) is re-plotted in logarithmic scale for clear display of the crossovers.

can be calculated to be at a distance $d_0 = -D_{2\theta}/D_\theta$ from the CoB. It was shown that at the CoH, both the coupled diffusion coefficients (D$_{1\theta}$, D$_{2\theta}$) are zero, D$_{11}$ remains the same as that at the CoB on the symmetric axis, and $D_{22}$ reaches its minimum, $D_{22}^{CoH} = 0.060 \ \mu m^2/s$.[19]

Based on the relation $D_{2\theta} = rD_\theta$, we can study Brownian motion at different levels of translation-rotation coupling by choosing different TPs (i.e. $r$) on the symmetry axis $X_1$ (Fig. 1). For the CoH, the translation-rotation coupling is zero in the body frame, $D_{2\theta} = 0 \ \mu m \cdot rad/s$. For the CoB, the translation-rotation coupling is $D_{2\theta} = -0.051 \ \mu m \cdot rad/s$. In addition, we choose a third point P on the opposite side of the CoH with respect to the CoH and with a $5d_0$ distance from the CoB (Fig. 1). The $D_{2\theta}$ at P is thus 5 times of that at CoB. The trajectories for the CoH and P were obtained from the trajectory of the CoB by simply shifting the instantaneous CoB positions by vectors of (d$_0$cosθ, d$_0$sinθ) and (-5d$_0$cosθ, -5d$_0$sinθ) respectively.

The MSDs calculated for these three tracking points are shown in Fig. 2. At the CoH, the MSDs grow linearly with time over the full range of observation time, similar to that of ellipsoids. For the CoB and P, the MSDs grow linearly with time at short and long times, while in the intermediate crossover regime the diffusion is sub-diffusive (non-Fickian). It can also be seen that the diffusion coefficients of CoB and P at long times are the same as the diffusion coefficient of the CoH (Fig. 2a).

This crossover behavior is agreements with the Langevin theory in which the MSD of a point on $X_1$ at a distance $r$ from CoH is given by:
$$\langle [\Delta x_{1,2}(x)]^2 \rangle = 2\overline{D}_{CoH}t + r^2(1 - e^{-D_\theta t}) \quad (2)$$
where $\overline{D}_{CoH} = (D_{11}^{CoH} + D_{22}^{CoH})/2$. This relationship yields the short time diffusion coefficient, $D_{TP}^S = \overline{D}_{CoH} + r^2 D_\theta/2$, and the long time diffusion coefficient $D_{TP}^L = \overline{D}_{CoH}$. Using the measured diffusion coefficients, we see that Eq. 2 agrees excellently with the experimental data (Fig. 2). The characteristic crossover time predicted by the Langevin theory is $t \sim 1/2D_\theta$, which, ~10s, is in agreements with the data. Since the second term of Eq. 2 is proportional to $r^2$, the crossover behavior for P is more distinguishable than that for the CoB.

To understand the translation rotational coupling effects when the CoH is not the TP, we measured the probability distribution functions (PDFs) $p(x_1, x_2, t)$ for displacements $x_1$ and $x_2$ after time $t$ in the lab frame with the initial orientations of the particle fixed at $\theta_0 = 0$ and the initial position of the tracking points at (0,0) (Fig. 3). Fig. 3a-d show the PDF at 4 different times with the CoH used as the tracking point. When the time increases, the PDF broadens in space due to diffusion while the maximal probability of finding the particle remains fixed at (0,0). In another word, the mean position of the CoH remains at the origin during the Brownian motion. This is the same as the case of ellipsoids where the translation and rotation are decoupled when the CoM is used for motion tracking. To note, for the short time such as $t = 1s$, the diffusion coefficients along $x_1$ and $x_2$ in the lab frame are the same as in the body frame. Since the diffusion coefficients $D_{11}^{CoH} = 0.049 \mu m^2/s$, $D_{22}^{CoH} = 0.060 \ \mu m^2/s$ are quite close, the anisotropic feature of the diffusion is only discernible in $p(x_1, x_2, t)$ when $t$ is short.

In Fig. 3e-h where the CoB is used for motion tracking, we can see that with the increase of the time, the PDF broadens in space as expected, and the PDF is more elongated along the $x_2$ axis as the $D_{22}$ is increased from $D_{22}^{CoH}$ by the amount of $r^2 D_\theta$. Meanwhile, it is clear that the maximal probability of finding the TP (i.e. CoB) is drifting towards the CoH over time. We can also see that at large times such as t = 100s, the most probable position of the CoB overlaps with the CoH.

The implications of translation-rotation coupling can be best understood when the point P is used for motion tracking in Fig. 3i-l. The translation-rotation coupling becomes more prominent than the previous two cases and the local behavior of the particle becomes distinct. It can be clearly seen from Fig.3i-j that as the point P diffuses the PDF forms a crescent like shape due to strong negative coupling between the translation along $x_2$ and rotation. Consequently, the average position of the TP P gradually moves along $x_1$ towards the CoH over time, and then the average position of the TP coincides with the CoH when the PDF forms a ring like shape encircling the CoH (Fig. 3k). This account for the non-zero MDs and the saturation of MDs at the distance between the TP and the CoH as observed for fixed initial orientation in our pervious results.[19] At long enough time the PDF ultimately becomes an isotropic dome shape with its peak located at the CoH and the motion of the TP and CoH becomes indistinguishable.

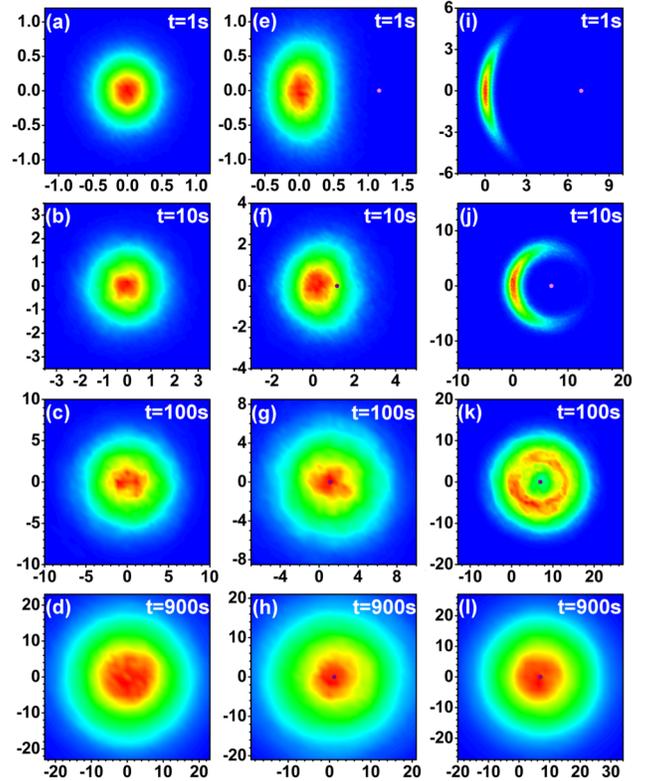

**Fig. 3**. The measured 2D probability distribution functions $p(x_1, x_2, t)$ at different times (1s, 10s, 100s, and 900s) for the CoH (a-d), the CoB (e-h), and P (i-l) as the TPs. The particle initial orientation is fixed at $\theta_0 = 0$. The horizontal and vertical axes are displacements along the $x_1$ and $x_2$ axes in the lab frame with units in μm. The dots in (e-l) indicate the position of the CoH.

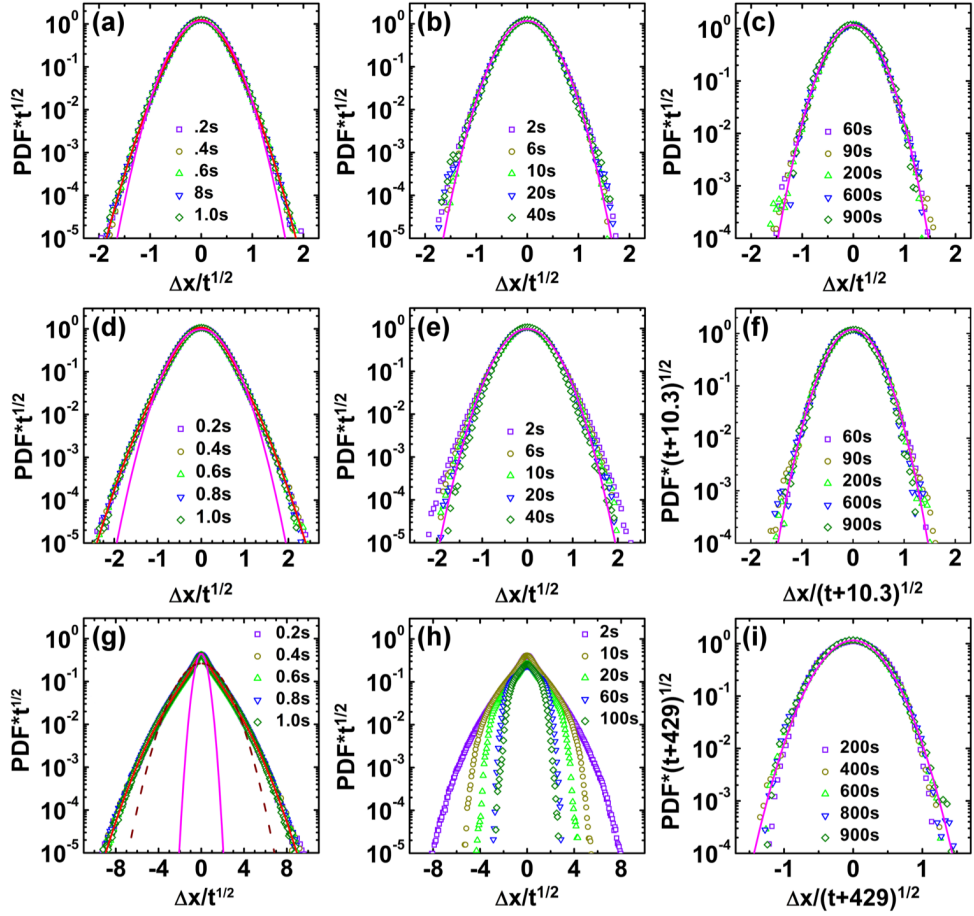

**Fig. 4** Scaling plots of the angle averaged PDFs in three different time regimes: short (0.2s-1s), intermediate (2s-40s) and long (60s-900s) times for the CoH (a-c), the CoB (d-f) and the point P (g-i) as the TP. The red curves in (d, g) represent empirical fitting with equations described in the text. In (a-c), the Gaussian curves use the same diffusion coefficient $D^{CoH} = 0.058 \mu m^2/s$ for the CoH. In (d, e), the magenta Gaussian curves use a short time diffusion coefficient $\bar{D}^S_{CoB} = 0.082 \mu m^2/s$ (d) and a long time diffusion coefficient $\bar{D}^L_{CoB} = 0.058 \mu m^2/s$ (e). In (g), the magenta Gaussian curve is a Gaussian fitting to the central part of the scaled PDF; the dashed brown curve is the Gaussian function with the short time diffusion coefficient $\bar{D}^S_P = 1.15 \mu m^2/s$. In (i) the magenta curve is the Gaussian function with the long time diffusion coefficient $\bar{D}^L_P = 0.058 \mu m^2/s$ for P.

The fundamental impacts of the TP on the behavior of Brownian motion can be seen from these 2D PDFs for fixed initial angles. When the TP is coincident with the CoH, the short time 2D PDFs are of $D_2$ symmetry as an ellipsoidal disk (Fig. 3a-3b). When the TP is not coincident with the CoH, the short time 2D PDFs for the boomerangs exhibit the same $D_1$ symmetry as their geometric shapes (Fig. 3e-3f, Fig. 3i-3j). In another word, the symmetry of the short time 2D PDFs reflect actually the symmetry of the particle shape with respect to the TP used when the TP does not overlap with the CoH.

The angle-averaged PDF is directly related to the MSDs measured in the lab frames. As expected, the angle averaged PDF is significantly affected by the translation-rotation coupling (Fig. 4). To facilitate comparisons with the Gaussian distribution, we scaled the vertical axis as $PDF \cdot t^{1/2}$ and the horizontal axis as $x/t^{1/2}$. As for the CoH as the TP, the scaling plots collapse all the $p(x,t)$ data into master curves for all three time regimes. A comparison with the Gaussian function clearly shows that the PDFs deviate from Gaussian in both short and intermediate time regimes, while agree with Gaussian well in the long time regime. This non-Gaussian behavior originates from the difference between $D_{11}$ and $D_{22}$, the same as the Brownian motion of an ellipsoid.[9] While previous studies indicate that the short time PDFs for the ellipsoidal particles are quite complex (no analytical formula), we found that the PDFs at short times can be well fitted with an empirical formula (red curve in Fig. 3) $y = C_1[1 - exp(-C_2 x^2)] exp(-C_3 x^2)/x^2$: which is Gaussian for small x and becomes Levy like for large $x$; $C_1$=0.22, $C_2$=6.25 and $C_3$= 2.5 are the fitting parameters.

As for the CoB as the TP, the scaling plots collapse the PDF data in the short and long time regimes, while the PDFs in the crossover regime do not follow the scaling. In comparison with the CoH as TP, the deviation of the PDFs from the Gaussian distribution in the short time regime is increased. In particular, the probability of large displacements is increased, while the probability of small displacements still obeys approximately the Gaussian distribution with the short time diffusion coefficient $D_{CoB}^S = 0.082\ \mu m^2/s$. Again, we found that the PDFs can be well fitted with the same formula as for the CoH: $y = C_1[1 - exp(-C_2 x^2)]exp(-C_3 x^2)/x^2$ where $C_1$=0.22, $C_2$=4.35 and $C_3$= 1.43 are fitting parameters. At the intermediate times only the central part of the PDF follows the master curve and the large $x$ tails clearly are not scalable a a result of the evolving PDF shape and time-dependent diffusion coefficients. The PDFs in the intermediate time regime evolve eventually to the Gaussian form as the MSD reverts back to Fickian in long time (Fig. 4f). For the scaling, since the MSDs in Eq. 2 have a constant term $r^2$ at long time, we multiplied the long time PDFs and normalized $x$ by $(t + r^2/2D_{CoB}^L)^{1/2}$, and could scale all data to a Gaussian curve with a variance corresponding to the long time diffusion coefficient $D_{CoB}^L = 0.058\ \mu m^2/s$ (Fig. 4f).

The effects of translation-rotation coupling on the displacement PDFs are made conspicuous for the TP P. As seen from the Fig. 4g-i, similar scaling is still applicable for the short and long time PDFs, indicating constant diffusion coefficients therein. However, the scaled short time PDFs deviates significantly from Gaussian; a comparison with the Gaussian distribution (dotted curve) corresponding to the short time diffusion coefficient shows that the behavior is distinctively different from that of the CoB and the central part that fits with Gaussian (magenta curve) is much smaller than that obtained from the short time diffusion coefficient (brown dashed curve). The contribution to the PDF is predominantly from the large displacement steps which noticeably belong to a non-Gaussian distribution. We also found that the scaled short-time PDFs can be well fitted with a similar formula but with a different exponent $y = C_1[1 - exp(-C_2|x|)]\ exp(-C_3 x^2)/|x|$: where $C_1$ = 0.22, $C_2$ = 2.13 and $C_3$ = 0.1 are fitting parameters. Similarly, the PDFs for the intermediate time range do not follow the scaling behavior with increased discrepancies, and the long-time scaling of PDFs of point P shows a Gaussian distribution.

## 4. Discussions

The large translation-rotation coupling and the corresponding PDFs exhibited by the tracking point P have some relevance to prior experimental observations by Wang et al.[8,10] Two systems studied in their work involve translation-rotation coupling. One system is colloidal beads attached on phospholipid bilayer microtubes by electrostatic attractions, where the concentration of the microtubes is so low that interactions between them can be ignored. The second system is liposomes diffusing in a nematic solution of F-actin filaments. The Brownian motion of the particles can be considered as the superposition of 1D Brownian motion on the microtubes and the Brownian motions of a TP (i.e. the particle position) on rod-shaped microtubes/filaments. When the TP is not at the middle of the microtubes/filaments, its Brownian motion is similar to the P point of the Boomerang particle, involving large translation-rotation coupling. In fact, the PDFs measured for liposomes diffusing in the nematic filaments exhibit quite similar to the short-time PDFs of the point P of the boomerang particles. It will be interesting to look further into how translation-rotation coupling of rod-shaped particles contribute to the non-Gaussian PDFs of the tagging particles.

In addition, it is interesting to note that the cross-over from short time diffusion to long time slow diffusion of single low symmetry particles exhibit some phenomenological similarities with the self-diffusions of colloids in finite concentrations. In the self-diffusion, a tagged particle is considered with respect to the rest particles; its Brownian motion undergoes three characteristic time regimes. In the short time regime $t < a^2/4D$, i.e. when the tagged particle diffuses distances smaller than its size, the configurations of its surrounding particles do not change much, and the MSDs grow linearly with time with a short time diffusion deviated from that of single particles due mainly to the hydrodynamic interactions with neighboring particles. In the intermediate crossover regime $t \sim a^2/4D$, the tagged particle is caged by its neighboring particles and the growth of its MSDs with time slows down and becomes non-linear. In long time region $t > a^2/4D$, the particle escapes the cage and the MSDs grow linearly again with time with a smaller long time diffusion coefficient. Similarly, the Brownian motion of tracking point P is firstly within a cage formed due to the large translation-rotation coupling, and then regular diffusion outside the cage (Fig. 3i-l). Certainly the underlying physics of low symmetry particle diffusion and self-diffusion is quite different.

## 5. Conclusions

In conclusion, we studied the effects of translation-rotation coupling on the probability distribution functions of the random displacements of the boomerang colloidal particles. We found that with the increase of translation-rotation coupling, the 2D PDFs with fixed initial angle transform from an elliptical shape to a crescent shape, which can provide a clear physics picture of the non-zero mean displacements. We also can show that the angle averaged PDFs can be scaled into master curves in short times and long times. The long time PDFs can be well fitted with Gaussian functions, while the shape of the short time PDFs vary with the translation-rotation coupling. These observations are relevant to various experimental situations such as the Brownian motions of quantum dots tagged to microtubules or nanorods and fluorescent dyes tagged on large macromolecules, and thus suggest that at certain time scales, the effects of the translation-rotation coupling on the displacement PDFs and diffusion coefficients need to be carefully taken into account.

## Acknowledgements

The authors acknowledge valuable discussions with Andrew Konya, Jonathan V. Selinger and Aleksandar Donev. The work was supported by a Farris Family Award and partially supported by NSF ECCS-0824175.